\documentclass[showpacs,aps,graphicx,twocolumn]{revtex4}
\usepackage{graphicx}

\begin{document}

\title{Eavesdropping on the "Ping-Pong" Quantum Communication Protocol Freely in a Noise Channel
\footnote{Published in \emph{Chinese Physics} \textbf{16} (2),
277-281 (2007).}}
\author{ Fu-Guo Deng$^{a)b)c)}$\footnote{ Email address: fgdeng@bnu.edu.cn}
 Xi-Han Li$^{a)b)}$, Chun-Yan Li$^{a)b)}$, Ping Zhou$^{a)b)}$
and Hong-Yu Zhou$^{a)b)c)}$}
\address{$^{a)}$ The Key Laboratory of Beam Technology and Material
Modification of Ministry of Education, Beijing Normal University,
Beijing 100875,
China\\
$^{b)}$ Institute of Low Energy Nuclear Physics, and Department of
Material Science and Engineering, Beijing Normal University,
Beijing 100875, China\\
$^{c)}$ Beijing Radiation Center, Beijing 100875, China}
\date{\today }

\begin{abstract}
We introduce an attack scheme for eavesdropping the ping-pong
quantum communication protocol proposed by Bostr$\ddot{o}$m and
Felbinger [Phys. Rev. Lett. \textbf{89}, 187902 (2002)] freely in a
noise channel. The vicious eavesdropper, Eve, intercepts and
measures the travel photon transmitted between the sender and the
receiver. Then she replaces the quantum signal with a multi-photon
signal in a same state, and measures the photons return with the
measuring basis with which Eve prepares the fake signal except for
one photon. This attack increase neither the quantum channel losses
nor the error rate in the sampling instances for eavesdropping
check. It works for eavesdropping the secret message transmitted
with the ping-pong protocol. Finally, we propose a way for improving
the security of the ping-pong protocol.

\end{abstract}


\maketitle

Quantum mechanics offers some unique capabilities for the processing
of information, such as quantum computation and quantum
communication \cite{book}. Quantum cryptography, one of the most
mature quantum techniques, provides a novel way for transmitting of
message securely. Since Bennett and Brassard proposed the original
quantum key distribution (QKD) protocol in 1984, a lot of works have
been focused on this topic, such as
\cite{Gisin,longliu,Hwang,CORE,BidQKD,ABC,delay,cpqkd,wl01,wl02,wl03}.
In QKD, the two parties, say the sender, Alice and the receiver,
Bob, can create a random binary string with quantum channel
unconditionally secure \cite{book}. The no-cloning theorem
\cite{nocloning} forbids any eavesdropper to eavesdropping an
unknown quantum state without disturbing it. In fact, QKD is secure
as the authorized users can find out the eavesdropping done by Eve
if she wants to steal the quantum information, and then they discard
the string, which does not reveal the secret message.

Recently, a novel concept, quantum secure direct communication
(QSDC),  was proposed and actively pursued
\cite{beige,bf,two-step,QOTP,QSDC,cai,yan,zhangzj,Gao,gaocp,Nguyen,mancpl,zhangsPRA,song,lixh}.
Also, it is extended for controlled teleportation \cite{yangj}. With
QSDC, the secret message is transmitted directly without first
creating a random key to encrypt it, which is different from QKD
whose object is just to establish a common random key between two
remote parties. As the secret message cannot be altered by the two
authorized users when it has been transmitted in the quantum
channel, the security of QSDC depends on the fact that Eve can only
get a random outcome if she monitors the line
\cite{two-step,QOTP,QSDC}. Moreover, Alice and Bob can detect the
eavesdropping if Eve monitors the quantum line before they code the
message on the quantum states. By far, almost all the existing QSDC
protocols can be attributed to one of the two types. The first one
are the QSDC protocols in which the secret message can be read out
directly without exchanging an additional classical information for
each qubit except for the sampling qubits for eavesdropping check,
such as those in
Refs.\cite{bf,two-step,QOTP,QSDC,cai,Nguyen,mancpl}. The other one
are those protocols in which each qubit can be read out by the
legitimate user after at least a bit of classical information is
exchanged \cite{beige,zhangzj,yan,Gao,gaocp}. An interesting feature
of the QSDC protocols in Refs. \cite{two-step,QOTP,QSDC} is that the
quantum states are transmitted in a quantum data block and the two
legitimate users can maintain its security with error correction and
quantum privacy in a noise channel.

The famous ping-pong QSDC protocol \cite{bf} proposed by
Bostr$\ddot{o}$m and Felbinger has been claimed to be secure for
establishing a random key and quasisecure for transmitting a plain
text message (a secret message) as Eve is able to gain a small
amount of message information before being detected \cite{bf}.
Recently, the ping-pong protocol is  proven insecure if the
 quantum channel losses are high enough
\cite{attack1,attack2,attack3} even for distributing a common random
key. Also it can be attacked without eavesdropping
\cite{attack4,attack5}. In this paper, we will show that the
ping-pong protocol can be eavesdropped freely if the error rate
introduced by the quantum channel noise is not zero, not requiring
that the loss of the quantum channel is high. Moreover, we introduce
a way for improving its security in a noise channel.

\begin{figure}[!h]
\begin{center}
\includegraphics[width=8cm,angle=0]{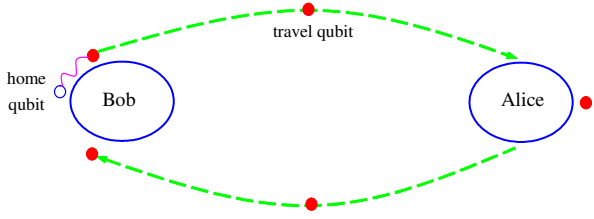} \label{f1}
\caption{ Schematic demonstration of the ping-pong QSDC protocol,
similar to the figure 1 in the Ref. \cite{QSDC}. Alice is the
sender of message, and Bob is the receiver.}
\end{center}
\end{figure}

In the ping-pong QSDC protocol \cite{bf}, the message receiver Bob
prepares the quantum source, an Einstein-Podolsky-Rosen (EPR) pair.
An EPR pair is in one of the four Bell states shown as following
\cite{book}:
\begin{eqnarray}
\vert \psi ^{\pm}\rangle =\frac{1}{\sqrt{2}}(\vert 0\rangle\vert
1\rangle\pm\vert
1\rangle \vert 0\rangle), \label{EPR12}\nonumber\\
\vert \phi ^{\pm}\rangle =\frac{1}{\sqrt{2}}(\vert 0\rangle\vert
0\rangle\pm\vert 1\rangle\vert 1\rangle), \label{EPR34}
\end{eqnarray}
where $\vert 0\rangle$ and $\vert 1\rangle$  are the horizontal and
vertical polarized states of a single photon, respectively. The two
photons in each EPR pair prepared by Bob are in the maximal
entangled state $\vert \psi ^{+}\rangle=\frac{1}{\sqrt{2}}(\vert
0\rangle_{H}\vert 1\rangle_{T} + \vert 1\rangle_{H} \vert
0\rangle_{T})$. Here $H$ and $T$ represent the home qubit and the
travel qubit \cite{bf}, respectively. Similar to quantum dense
coding \cite{densecoding}, Bob keeps the qubit $H$ and sends the
qubit $T$ to Alice. Alice chooses two modes, the control mode and
the message mode, for dealing with the $T$ qubit, i.e., a
probability $c$ for picking up the control mode for the photon and
$1-c$ for coding the message. When she chooses the control mode,
Alice performs a single-photon measurement on the $T$ qubit with the
horizontal-vertical measuring basis (MB), say $\sigma_z$, otherwise
she codes the photon with $I=\vert 0\rangle\langle 0\vert + \vert
1\rangle\langle 1\vert$ and $Z=\vert 0\rangle\langle 0\vert - \vert
1\rangle\langle 1\vert$ when the messages are 0 and 1, respectively.
\begin{eqnarray}
(Z\otimes I)\vert \psi ^{+}\rangle=\vert \psi ^{-}\rangle.
\end{eqnarray}

Cai introduced an attack way without eavesdropping \cite{cai}. In
Cai eavesdropping scheme, Eve measures the $T$ photon with the MB
$\sigma_z$. This attack cannot be detected if Alice and Bob only
take the MB $\sigma_z$ on their sampling photons. In an ideal
channel without noise and loss, this attack cannot get the
information about the secret message. However, we have to confess
that there are noises in a practical quantum channel which will
introduce an error rate $\varepsilon_c$ in the outcomes
\cite{book,Gisin}. With the improvement of technology,
$\varepsilon_c$ can be small, but not zero. Moreover, a
single-photon detector has a special recovery time (i.e., the dead
time) \cite{Gisin} in which the $N$ photons attained are recorded as
just one. Eve can exploit the error rate $\varepsilon_c$ and the
recovery time to hide her eavesdropping on the ping-pong protocol
and get almost all the information about the secret message with a
multi-photon fake signal even though the quantum channel loss is
low. We introduce it in detail as following, similar to the Trojan
horse attack in Ref. \cite{Gisin,dengattack}.

\begin{figure}[!h]
\bigskip
\begin{center}
\includegraphics[width=5cm,angle=0]{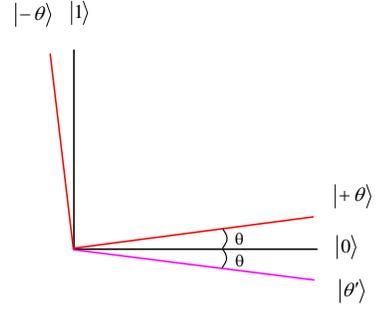} \label{f2}
\caption{ The state of the multi-photon fake signal. $\vert
+\theta \rangle$ and $\vert -\theta \rangle$ are the two
eigenstates of the measuring basis $\sigma_{\theta}$. }
\end{center}
\end{figure}

\begin{figure}[!h]
\bigskip
\begin{center}
\includegraphics[width=8cm,angle=0]{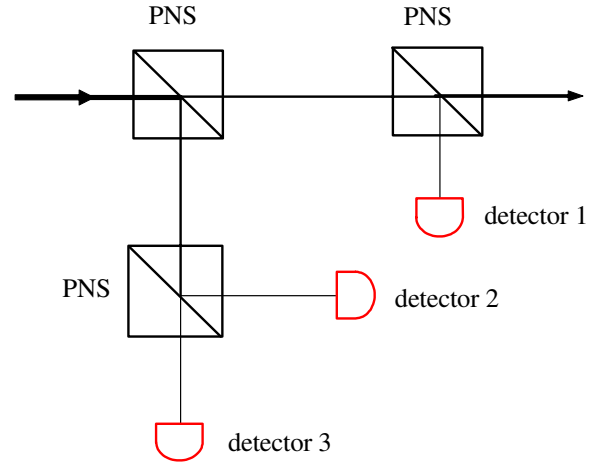} \label{f3}
\caption{ The attack with the photon number splitters (PNS: 50/50)
in the case that there are four photons in each fake signal. }
\end{center}
\end{figure}

For the eavesdropping, Eve first intercepts and measures the $T$
photon with MB $\sigma_z$, and then she prepares an $N$-photon fake
signal with the MB $\sigma_\theta$ whose two eigenstates can be
written as
\begin{eqnarray}
\vert +\theta\rangle&=&cos\theta \vert 0\rangle + sin\theta \vert
1\rangle,\nonumber\\
\vert -\theta\rangle&=&-sin\theta \vert 0\rangle + cos\theta \vert
1\rangle,
\end{eqnarray}
where $\theta\in [0, \frac{\pi}{2})$ and
\begin{eqnarray}
 sin^2\theta \leq \varepsilon_{c}.
\end{eqnarray}
When the outcome of the measurement is $\vert 0\rangle_{T}$, Eve
prepares the $N$-photon fake signal in the same state $\vert
+\theta\rangle=cos\theta \vert 0\rangle + sin\theta \vert 1\rangle$,
shown in Fig.2, and resends it to Alice in a time slot, shorter than
the recovery time of the single-photon detector. As its dead time,
Alice's detector only records a single photon when Alice measures
the signal by choosing the control mode with the MB $\sigma_z$. In
this way, Eve's eavesdropping will introduce the error rate
$\varepsilon_{E}= sin^2\theta$ in the sampling instances between
Alice and Bob. Eve can use a better quantum channel with which the
error rate is lower by far than the origin one to hide her
eavesdropping freely.

As an example for demonstrating the principle of this attack, we
assume that  $\varepsilon_{c}=10\%$ and Eve uses an ideal quantum
channel to steal the message below. As the symmetric, we assume
that there are $N=2^m$ photons in the fake signal.
\begin{eqnarray}
 \varepsilon_{E}=sin^2\theta = \varepsilon_{c}=0.1.
\end{eqnarray}

After the coding done by Bob with one of the two local unitary
operations $I$ and $Z$, Eve intercepts the fake signal again. She
splits the multi-photon signal with some photon number splitters
(PNS: 50/50), and sends one photon to Bob and measures the other
photons, see in Fig.3.

If Alice performs the $I$ operation on the fake signal, the
photons in the fake signal are in the state $\vert T'\rangle=\vert
+\theta\rangle= cos\theta \vert 0\rangle + sin\theta \vert
1\rangle$; otherwise $\vert T'\rangle=\vert \theta'\rangle=
cos\theta \vert 0\rangle - sin\theta \vert 1\rangle$. The attack
for obtaining the information about the local unitary operations
done by Alice is simplified to distinguish those two states. It is
impossible for Eve to get almost all the information about Alice's
operation if she has only one photon coded by Alice as $\vert
\langle \theta' \vert +\theta\rangle \vert^2=cos^22\theta=0.64$.
But the story is changed if there are many photons in each fake
signal. Eve can distinguish those two states with a large
probability and then steal almost all of the message freely.

Fig.3 gives us an example for Eve to eavesdrop the message with four
photons in each fake signal. Eve splits the fake signal with three
PNS when the signal returns from Alice to Bob. She sends one of the
four photons to Bob and measures the other three photons with the MB
$\sigma_{\theta}$, see Fig.2. If the three photons are all in state
$\vert +\theta\rangle= cos\theta \vert 0\rangle + sin\theta \vert
1\rangle$, i.e., Alice performs the identity operation $I$ on the
fake signal, Eve gets the outcome $\vert +\theta\rangle$ with the
probability 100\%; otherwise Eve has the probability
$(cos^22\theta)^{n-1}=(0.64)^3=0.262144$ to obtain the state $\vert
+\theta\rangle$ for her measurements on all the three photons. That
is, Eve has the probability $P_F=0.262144$ that she will fail to
distinguish the two operations done by Alice on the fake signal. If
there are $N$ photons with which Eve distinguish the two states
$\vert +\theta\rangle$ and $\vert \theta'\rangle$, the probability
that Eve will fail is reduced to
$P_F=(cos^22\theta)^{n-1}=(0.64)^{n-1}$. When n=64, $P_F\cong
6.16\times 10^{-13}$. It means that Eve can obtain the message fully
if there are a large number of photons in each fake signal as this
attack increases neither  the signal losses nor the error rate in
the sampling instances.

In essence, the security issue in ping-pong QSDC protocol \cite{bf}
arose from the fact that the two authorized users transmit the
qubits one by one and check the eavesdropping only with the same MB
$\sigma_z$. The secret message transmitted cannot be discarded,
different from the outcomes in QKD \cite{Gisin}.  For improving its
security, it is necessary for Alice and Bob to transmit the qubits
in a quantum data block, similar to \cite{two-step,QOTP,QSDC}, and
measure the sampling instances with two MBs $\sigma_z$ and
$\sigma_x$. Here $\sigma_x = \{ \frac{1}{\sqrt{2}}(\vert 0\rangle +
\vert 1\rangle), \frac{1}{\sqrt{2}}(\vert 0\rangle - \vert
1\rangle)\}$. As the eavesdropping check depends on the public
statistical analysis of the sampling instances, the transmission of
the quantum data block ensures that the message is coded after the
verification process is accomplished. Moreover, the two parties can
do quantum privacy amplification on the quantum date
\cite{two-step,QOTP,QSDC} before Alice codes her message on the
quantum states. Those two interesting characters paly an important
role in the security of QSDC protocols.

With the two MBs for the sampling instances, the action done by the
eavesdropper, Eve will leave a trace in the results and will be
detected. Moreover, this modification can improve the capacity in
the ping-pong QSDC protocol, as discussed in Ref. \cite{caiA}. For
most of the existing QSDC protocols, there is a probability that Eve
can get a part of message if she eavesdrops the quantum channel with
a Trojan horse attack strategy \cite{Gisin} and replacing the
original quantum channel with an ideal one. In the QSDC protocols
\cite{two-step,QOTP,QSDC}, the parties can reduce the information
leaked to Eve to a negligible value with quantum privacy
amplification \cite{Gisin,two-step,QOTP}. Also, Alice and Bob can
prevent Eve from eavesdropping with this attack if they use some PNS
to monitor the sampling instances. That is, they split the signal
with some PNS and measure them individually with choosing the MB
$\sigma_z$ and $\sigma_x$ randomly, similar to Ref.
\cite{dengattack}. This strategy for eavesdropping check can also be
used to improve the security in the QKD protocol \cite{BidQKD} and
the secure deterministic communication protocol \cite{Lucamaini}
proposed by Lucamarini and Mancini following the ideas in Refs.
\cite{BidQKD,QOTP}. In a practical application, the users can also
use some photon beam splitters to replace the PNSs.

In conclusion we have presented an attack strategy on the ping-pong
QSDC protocol freely in a noise quantum channel. This attack works
for getting the secret message transmitted with the ping-pong
protocol \cite{bf}. The eavesdropper, Eve can intercept the signal
transmitted between Alice and Bob and measures it first, and then
she replaces it with a multi-photon fake signal. Eve's eavesdropping
can be hidden by the error rate introduced by the noise in the
practical quantum channel and the dead time of a detector. She can
obtain almost all the information about the message with some photon
number splitters and measurements along some a direction. We also
suggest the way for improving the security of the ping-pong protocol
and introduce a way for prevent the eavesdropper from stealing the
information with the Trojan horse attack strategy.

This work was supported by the National Natural Science Foundation
of China under Grant No. 10604008 and Beijing Education Committee
under Grant No. XK100270454.

\end{document}